\documentclass[namedreferences]{solarphysics}
\usepackage[hyperref,optionalrh,natbib]{spr-sola-addons}
\usepackage{cite}
\usepackage{graphicx}
\usepackage{color}
\def\we{Waldmeier effect}
\def\sc{solar cycle}
\def\amp{amplitude}

\def\mc{meridional circulation}

\def\ftdm{flux transport dynamo model}
\def\Rs{R_{\odot}}
\def\er{\mbox{erf}}
\def\blue{\textcolor{black}}
\newcommand{\Fig}[1]{Figure~\ref{#1}}

\begin{document}
\begin{article}
\begin{opening}

\title{Correlation between Decay Rate and Amplitude of Solar Cycles as Revealed from Observations and Dynamo Theory}
\author{\surname{Gopal Hazra}$^{1,3}$\sep
            ~\surname{Bidya Binay Karak}$^{2}$\sep
            ~\surname{Dipankar Banerjee}$^{3}$\sep
           ~\surname{Arnab Rai Choudhuri}$^{1}$}

\institute{$^{1}$~Department~of~Physics,~Indian~Institute~of~Science,~Bangalore~560012,~India\\
                     email: \url{ghazra@physics.iisc.ernet.in}\\ email: \url{arnab@physics.iisc.ernet.in}\\ 
             $^{2}$~NORDITA~KTH~Royal~Institute~of~Technology~and~Stockholm~University,\\
                    Roslagstullsbacken~23,~SE-106~91~Stockholm,~Sweden\\
                     email: \url{bbkarak@nordita.org} \\
           $^{3}$~Indian~Institute~of~Astrophysics,~Bangalore~560034,~India\\
                     email: \url{dipu@iiap.res.in}\\
             }

\runningtitle{Correlation of the Amplitude of the Solar Cycle with Decay Rate and Period}
\runningauthor{Hazra \it{et al.}}

\begin{abstract}
Using different proxies of solar activity, we have studied the following features of
solar cycle. ({i}) A linear correlation between the amplitude of cycle and its decay
rate, ({ii}) a linear correlation between the amplitude of cycle $n$ and the decay 
rate of cycle $(n - 1)$ and ({iii}) an anti-correlation between the amplitude of cycle 
$n$ and the period of cycle $(n - 1)$. Features ({ii}) and ({iii}) are very useful because 
they provide precursors for future cycles. We have reproduced 
these features using a flux transport dynamo model with stochastic fluctuations in the
Babcock-Leighton $\alpha$ effect and in the meridional circulation.  
Only when we introduce fluctuations in meridional circulation, we are able to 
reproduce different observed features of solar cycle. We discuss the possible reasons 
for these correlations.
\end{abstract}
\end{opening}

\section{Introduction}
Solar cycles are  asymmetric with respect to their maxima, the rise time being
shorter than the decay time. While the cycle amplitude (peak value) and the duration have cycle-to-cycle variations, we find some correlations among different
quantities connected with the solar cycle. Since 1935, it has been realized 
that the stronger cycles take less time to rise than the weaker ones 
(Waldmeier, 1935). This anti-correlation between
rise times and peak values of the solar cycle is popularly known as the \we.
\citet{KarakChou11} have defined this aspect of the \we\ as WE1,
whereas the correlation between the rise rates and the peak values is called WE2
(see also \opencite{CS08}).
Although WE2 is a more robust feature of the solar cycle,
\citet{KarakChou11} have shown
that  both WE1 and WE2 exist in many proxies of the solar cycle.
WE2 provides a valuable precursor for predicting solar cycles because one can predict the strength of
a cycle once it has just started (see \opencite{Lantos00,Kane08} ).

The declining phase of the cycle also provides important clues for understanding long-term variations.
We find that stronger cycles not only rise rapidly but also fall rapidly (shorter decay time). 
This results in a good correlation between the decay rate and amplitude of the same cycle.
However, defining the decay rate differently, \citet{CS08}
did not find a significant correlation between the decay rate and amplitude.
Furthermore, we find a strong correlation between the decay rate of the current cycle and 
the amplitude of the next cycle, 
\blue{which was also found by Yoshida and Yamagishi (2010).} 
The decay time, however, is found to 
have no correlation with the amplitude of the same cycle. 
Another important feature observed is that the amplitude of the cycle
is inversely correlated with the period of the previous cycle \blue{\citep{Hathaway02, Solanki02, Ogurtsov11}}. 
These two correlations again provide promising precursors 
to predict the strength of the future
cycle \citep{Solanki02, Watari08}.

Apart from showing these correlations from observational data, we also attempt
to provide theoretical explanations for them. A dynamo mechanism operating in the solar convection zone is believed to be 
responsible for producing the solar cycle. 
It is generally accepted that the strong toroidal field (responsible for the
formation of bipolar sunspots) is produced from the poloidal field by differential rotation
in the solar convection zone \citep{Parker55a}. This is the first part of solar dynamo theory. Due
to magnetic buoyancy \citep{Parker55b} the flux tubes of toroidal field erupt out
through the surface to form bipolar sunspot regions. These bipolar sunspots
acquire tilts due to the action of the Coriolis force during their journey through
the convection zone, giving rise to Joy's law \citep{Dsilva93}. To
complete dynamo action, the toroidal field has to be converted back into the
poloidal field. One possible mechanism for generating the poloidal
field is the Babcock--Leighton (B-L) process \citep{Bab61,Leighton69}, for which we now have strong observational support \citep{DasiEspuig10, Kitchatinov11a, Munoz13}. 
In this process, the fluxes of tilted bipolar active regions spread on the solar surface through
different processes (diffusion, meridional circulation, differential rotation)
to produce the poloidal field.
A model of the solar dynamo that includes a coherent meridional circulation and this B-L mechanism for
the generation of the poloidal field is called the flux transport dynamo model. This model  was proposed in the 1990s
\citep{WSN91,Durney95,CSD95}
and has been successful in reproducing many observed regular
as well as irregular features of the solar cycle \citep{CD2000,Kuker01, Nandy02, CNC04, Guerrero04, CK09, Hotta10, KarakChou13}. Recently \citet{Charbonneau10}
, \citet{Chou11} and \citet{Karakreview14} have reviewed this dynamo model. 

An important ingredient in flux transport dynamo is the \mc, which is not completely constrained 
either from observations or from theoretical studies.
Until recently not much was known about the detailed structure of the meridional
circulation in the convection zone \citep{Zhao13, Schad13}. Therefore, most of the
dynamo models use a single-cell \mc\ in each hemisphere.
However, very recently \citet{HKC14} have shown that a complicated multi-cellular \mc\
also retains many of the attractive features of the flux transport dynamo model if there is an equator-ward
propagating meridional circulation near the bottom of the convection zone or if
there is an equator-ward turbulent pumping \citep{Guerrero08}.
\blue{While most of the calculations in this paper are done for a single-cell
meridional circulation, we show that the results remain qualitatively similar
for more complicated meridional circulations.}

Since we want to do a theoretical study of the irregularities in the solar cycle, 
let us consider the sources of irregularities in the \ftdm\  that make different solar cycles unequal.
At present we know two major sources: (i) variations in the poloidal field generation due
to fluctuations in the B-L process \citep{CCJ07,GoelChou09} and (ii) variations in the meridional circulation \citep{Karak10, KarakChou11}. 
Direct observations of the polar field during last three cycles 
\citep{SCK05}, as well as its proxies such as the polar faculae and the active network index available for 
about last 100 years \citep{Munoz13,Priyal14}, indicate  large cycle-to-cycle variations of the polar field. 
The poloidal field generation mechanism mainly depends 
on the tilts of active regions, their magnetic fluxes and the meridional circulation,
all of which have temporal variations. Particularly the scatter of tilt angles
around the mean, caused by the effect of convective turbulence on rising flux
tubes \citep{Longcope02}, has
been studied by many authors \citep{WS89,DasiEspuig10}. 
Recently \citet{Jiang14} 
found that the tilt angle scatter led to a variation in the polar field  by
about $30\%$ for cycle 17. In fact, even a single big sunspot group with large tilt angle 
and large area appearing 
near the equator can change the polar field significantly \citep{Cameron13}.
On the other hand, for the meridional circulation, we have some surface measurements for 
about last 20 years, showing significant temporal variations (Chou and Dai, 2001; Hathaway and Rightmire 2010). 
Although our theoretical understanding of the \mc\ is very limited, a
few existing spherical global convection simulations do show significant variations
in the \mc\ \citep{PCB12,Karak15}.
Introducing randomness in the poloidal field generation and in the 
\mc, Karak and Choudhuri (2011) have been able to reproduce the Waldmeier effect in their
high diffusivity dynamo model.
When the \mc\ becomes weaker, the cycle period and hence the rise time becomes longer.
The longer cycle period allows the
turbulent diffusion to act for a longer time, making the cycle amplitude weaker
\citep{Yeates08,Karak10} and leading to the Waldmeier effect.
The variation of the meridional circulation is crucial in 
reproducing this effect.

The motivation of the present work is to explore how
the decay rates of cycles are related to their amplitudes in a flux transport dynamo model, with the aim of explaining the observed correlations
mentioned earlier.
The presentation of the paper is following. In the next section, we summarize
some of the features of solar cycle that are often considered as precursors
of the solar cycle. In Section~3,
we present a brief summary of our flux transport dynamo model and
then in Sections~4 we introduce suitable stochastic fluctuations in the poloidal
field and the meridional circulation, in order to reproduce various
observed features of the solar cycle. 
Finally the last section summarizes our conclusions.


\section{Observational Studies}
We have used three different observational data sets: ({i})
Wolf sunspot number{\footnote{http://solarscience.msfc.nasa.gov/greenwch/spot\_num.txt}} 
(cycles 1--23),
({ii}) sunspot area{\footnote{http://solarscience.msfc.nasa.gov/greenwch/sunspot\_area.txt}} 
(cycles 12--23), and ({iii}) $10.7$~cm radio flux{\footnote{http://www.ngdc.noaa.gov/stp/solar/flux.html}} (available only for the last five cycles).
These parameters are very good proxies of magnetic activity and are often used to study the solar cycle (Hathaway {\it {et al.}}\ 2002).
To minimize the noise while keeping the underlying properties unchanged, 
we smooth these monthly data using a Gaussian filter having a full-width at half maximum (FWHM) of 
1~year. We also average the data with FWHM of 2~years to check how the results change with
the filtering.

\begin{figure}
\centerline{\includegraphics[width=1.0\textwidth,clip=]{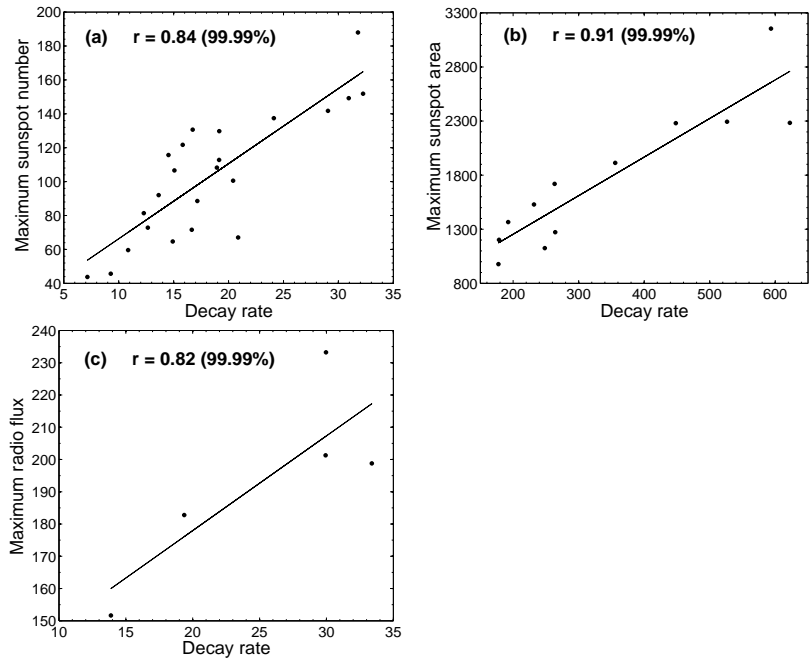}  } 
\caption{Scatter plots of the decay rate and the amplitude of the same cycle computed 
from (a) sunspot number, (b) sunspot area, and (c) $10.7$~cm radio flux data. 
In all these cases the original monthly data are smoothed using a Gaussian filter with FWHM of 2~years.The straight line in each plot is best linear fit of data. The correlation
coefficients ($r$) and the significance levels are also given in each plot.}
\label{obs1}
\end{figure}

\subsection{Correlation between the Decay Rate and the Cycle Amplitude}
We have calculated the decay rate at three different phases of the descending phase of the cycle, 
namely, early phase, late phase and entire phase. 
For the early phase, the decay rate is taken as the slope between two points with a separation of 1~year
with the first point one year after the cycle peak, whereas for the late phase the second point 
is taken 1~year before the cycle minimum. 
Here we exclude one year after the maximum when computing decay rate for the early phase
because sometimes the cycle peaks are not so prominent. While computing the decay rate for 
the late phase we also exclude 1~year before the minimum just to avoid the effect of 
overlapping between two cycles during solar minimum.
Finally, the decay rate of the entire decay part ({\it i.e}, entire phase) is taken as the average of the individual
decay rates computed at four different locations with a separation of one year starting from early
phase to the late phase. In \Fig{obs1}~(a), (b) and (c), we show the correlations of the cycle amplitudes with 
the decay rates of the entire phase computed from sunspot number, sunspot area and 
$10.7$~cm radio flux data, respectively.

We would like to point out that Cameron and Sch\"ussler (2008) have computed the decay rate 
from the intervals of two fixed values of solar activity and they did not get 
significant correlation between the decay rate and the amplitude  
(see right column, Figure\ 2 of their paper).
The reason of not finding significant correlation is that they have calculated 
the decay rate in the late phase of the cycle, {\it i.e.} near the tail of the cycle 
where the rate of decay is really very small. 
We find that their values are  comparable with our decay rates computed in the late phase.
In 4th and 5th columns of Table~1 we have listed our values and the values computed following 
Cameron and Sch\"ussler (2008) method (hereafter referred as CS08).
It is interesting to note that even for the radio flux data for which we have only five
data points, we get strong correlation; see Table~1 for details. 
Therefore we can see that if we determine the 
decay rates from the entire phase of the solar 
cycle or the early phase, we find strong correlation with the amplitude. 
Thus, to determine the decay rate from descending part of the solar cycle, we
need to consider the entire decay phase of the cycle, which provides a better estimate than CS08. 
\begin{table}
\caption{Correlation coefficients between different quantities of the solar cycle.}
\begin{tabular}{ccccccccc}
\hline
&&\multicolumn{6}{c}{Correlation coefficients of the decay rate with}& Correlation \\
&&\multicolumn{6}{c}{the amplitude of}& between the \\
\cline{3-8}
&&\multicolumn{4}{c}{Same cycle}&\multicolumn{2}{c}{Next cycle}& amplitude \\
\cline{3-8}
&&Entire &\multicolumn{2}{c}{Late decay phase}& Early & Entire & Late & and the\\
\cline{4-5}
Data set&FWHM &phase&Our & CS08's &Phase&phase&phase & previous\\
&&& value & value&&& & cycle period \\
\hline
Sunspot&1~yr&0.79&0.21&0.22&0.67 & 0.55 & 0.61 & -0.64\\
number&2~yr&0.86&0.45&--&0.86 & 0.65 & 0.83 & -0.67 \\ 
\hline
Sunspot&1~yr&0.84&0.20&0.11&0.69 & 0.14 & 0.37 & -0.49\\
area&2~yr&0.91&0.53&--&0.92 & 0.39 & 0.66 & -0.60\\ 
\hline
Radio&1~yr&0.86&-0.11&0.14&0.93&-0.42 & 0.64 & 0.11\\
flux &2~yr&0.82&0.24&--&0.95 & -0.43& 0.46 & 0.09 \\ 
\hline
\end{tabular}
\end{table}

\begin{figure}[!h]   
\centerline{\includegraphics[width=1.0\textwidth,clip=]{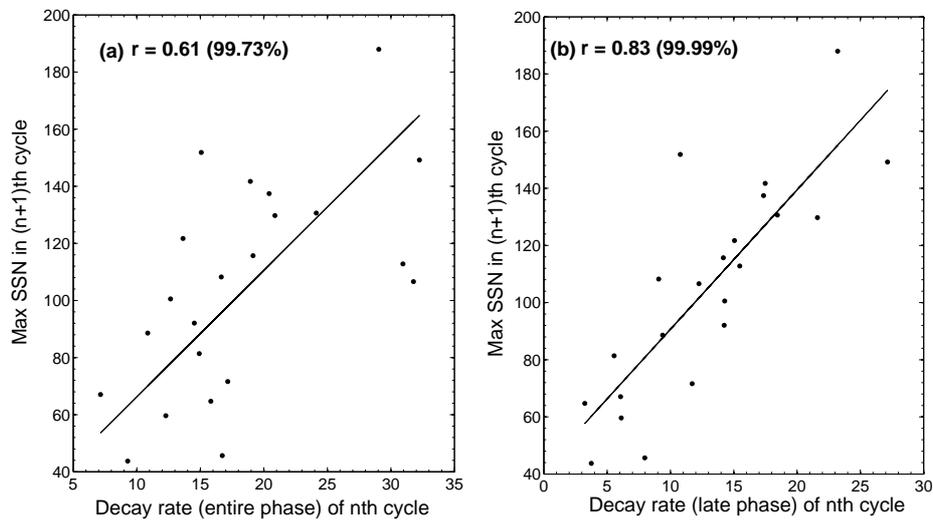} }
\caption{ Scatter plots showing the correlation of the amplitude vs. 
the decay rate of the previous cycle computed from sunspot number data (smoothed with FWHM of 2~year).
In (a) the decay rate is computed from the entire decay phase, whereas in (b) it is at 
late decay phase.} 
\end{figure}

\subsection{Correlation between the Decay Rate and the Next Cycle Amplitude}
Next we find that there is a significant correlation between the amplitude of 
cycle and the decay rate of the previous cycle. Again we find this correlation for all the data sets
considered here (see Table~1). However in Figure~2(a) we show this correlation only 
for sunspot number. Note that here the decay rates have been calculated from 
the entire decay phase as discussed in Section 2.1.
This correlation suggests that the decay rate of a cycle carries some information 
of the strength of the next cycle.
It is interesting to note that when we look at this correlation with the 
decay rate computed in the late phase, the correlations become even stronger; see Figure~2(b).
In 7th and 8th column of Table~2, we show both correlations for all three data sets. 
These results suggest that particularly the late phase of the cycle carries more 
information of the forthcoming cycle.
\blue{This correlation of decay rate with the amplitude of succeeding cycle is already reported in {\citet{Yoshida10}}. They have shown
this correlation for only sunspot number data and their methodology for calculation of 
decay rate (rate of decrease in sunspot number over some time) is somewhat different from our methodology.
They have studied decay rate in six different cases (see their Figures 1(a)-(f)). They have obtained the decay rate from
the decrease of sunspot number (SSN) over the period of 1, 2, 3, 4, 5 and 6 years before the minima 
of the cycle in the six different cases of study respectively.  Since solar cycles 
sometimes have overlapping regions during minima and it is difficult to ascertain
the actual minima, there are some uncertainties in the methodology of Yoshida
and Yamagishi (2010). 
The correlation coefficient ($r$ = 0.70) obtained in the second case of their study (see their Figure 1(b)) 
should be the same with what we obtained during late phase correlation ($r$ = 0.83). Since they have not considered the overlapping region between the minima and used monthly smoothed SSN, the value of the correlation coefficient is slightly different.}

\citet{CS07} (also see \opencite{Brown76}) have observed similar feature that the 
activity level during the solar minimum is an indicator for the strength 
of the next solar cycle and argued that this is caused by overlap between 
two cycles during solar minimum.

In all our theoretical calculations (subsequent section), while studying the correlation between the 
amplitude and the decay rate of the same cycle, we shall consider the decay rate of the 
entire phase, but for the correlation with the next cycle we shall consider only 
the late-phase decay rate.

\begin{figure}[!h]
\centerline{\includegraphics[width=0.8\textwidth,clip=]{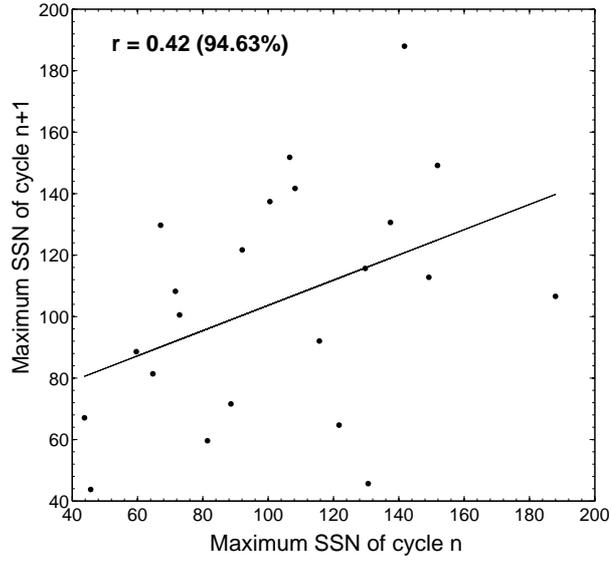}} 
\caption{ Scatter plot of $n$th cycle amplitude 
and the amplitude of the next $n+1$ cycle from sunspot number data (smoothed with FWHM of 2~years).}
\label{amplcorl}
\end{figure}

Since the decay rate of the cycle $n$ is correlated both with the amplitude
of cycle $n$ (Figure~1) and the amplitude of cycle $n+1$ (Figure~2), one question that
naturally arises is whether the amplitude of cycle $n$ and the amplitude 
of cycle $n+1$ are themselves correlated.  We show a correlation plot
between these amplitudes in Figure~3, demonstrating that there is not a
significant correlation. The challenge before a theoretical model is, therefore,
to explain how the decay rate of cycle $n$ is correlated both with the amplitude
of cycle $n$ and the amplitude of cycle $n+1$, while these amplitudes themselves
do not have a strong correlation.

\subsection{Correlation between the Cycle Period and the Next Cycle Amplitude}
\begin{figure}[!h]
\centerline{\includegraphics[width=0.75\textwidth,clip=]{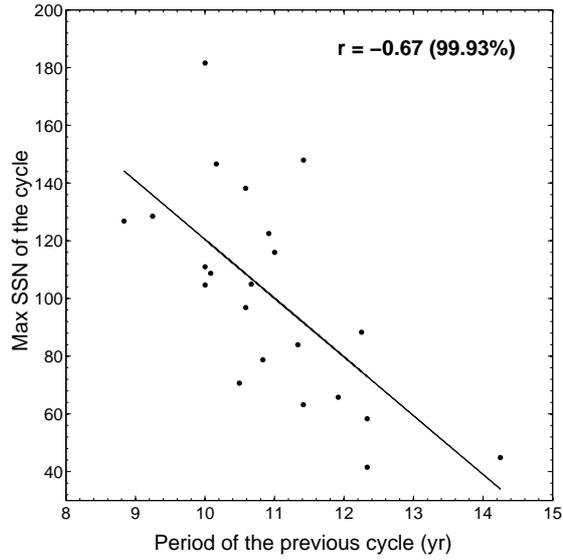}} 
\caption{Scatter plot showing the anti-correlation between the cycle amplitude 
and the period of the previous cycle from sunspot number data (smoothed with FWHM of 2~years).}
\label{percorl}
\end{figure}

Finally, we also find that the shorter cycles are followed by stronger cycles and 
vice versa. This produces an anti-correlation between the amplitude of a cycle 
and the period of the previous cycle \blue{\citep{Hathaway02, Solanki02, Ogurtsov11}}. \Fig{percorl} shows this correlation from 
sunspot number data (smoothed using a Gaussian filter with FWHM of 2~years). 
The correlation coefficients  from other data are listed in Table~1. For all data we have taken the period of 
the cycle just as the time difference between two successive minima.

\section{Theoretical Framework of the Dynamo Model}
\label{sec:model}
We carry out our theoretical studies using the flux transport dynamo model
originally presented by Chatterjee, Nandy, and Choudhuri (2004). In this model, the evolution of 
the axisymmetric two-dimensional magnetic field is governed by following two equations:
\begin{equation}
\frac{\partial A}{\partial t} + \frac{1}{s}({\bf v}.\nabla)(s A)
= \eta_{\rm{p}} \left( \nabla^2 - \frac{1}{s^2} \right) A + S_{\rm{BL}}(r,\theta;B),
\label{eqA}
\end{equation}
\begin{equation}
\frac{\partial B}{\partial t}
+ \frac{1}{r} \left[ \frac{\partial}{\partial r}
(r v_r B) + \frac{\partial}{\partial \theta}(v_{\theta} B) \right]
= \eta_{\rm{t}} \left( \nabla^2 - \frac{1}{s^2} \right) B 
+ s({\bf B}_{\rm{p}}.\nabla)\Omega + \frac{1}{r}\frac{d\eta_{\rm t}}{dr}\frac{\partial{B}}{\partial{r}},
\end{equation}
where $s = r \sin \theta$, $B (r, \theta)$ is the toroidal component of the magnetic field , 
$A(r, \theta)$ is the vector potential of the poloidal field,
${\bf v}=v_r{\bf \hat r} + v_\theta\hat {\bf \theta}$ is the velocity of the meridional flow, 
$\Omega$ is the internal angular velocity of the Sun and $\eta_{\rm{t}}$, $\eta_{\rm{p}}$ are the
turbulent diffusivities of the toroidal and the poloidal fields. Since the detailed discussion of the 
parameters and boundary conditions are given in Chatterjee, Nandy, and Choudhuri (2004) and 
Karak and Choudhuri (2011), here we do not discuss them again. 
We only make a few remarks about magnetic buoyancy and about the term $S_{\rm{BL}}(r,\theta;B)$ appearing in Equation (\ref{eqA}), which captures the longitude averaged B-L
mechanism. 

Let us discuss how the magnetic buoyancy is treated in this model.
When the toroidal field above the tachocline ($r= 0.71 \Rs$) at any latitude exceeds a certain
value, a fraction of it is reduced there and the equivalent amount of this field is added on the solar 
surface. Then this local toroidal field near the surface is multiplied by a factor $\alpha$ to 
give the poloidal field. The source term in Equation~(\ref{eqA}), therefore, is
\begin{equation}
S_{\rm{BL}}(r,\theta;B)=\alpha B(r,\theta,t),
\label{alphaH}
\end{equation}
where
\begin{equation}
\alpha =\frac{\alpha_0}{4} \cos \theta 
\left[ 1 + \er \left(\frac{r - 0.95\Rs}{0.03\Rs} \right) \right]
\left[ 1 - \er \left(\frac{r - \Rs}{0.03\Rs} \right) \right],
\end{equation}
with $\alpha_0 = 30$~m~s$^{-1}$.
Now our job is to use this model to study the observed features of solar cycle reported in previous sections.
To study any irregular feature of the solar cycle, we have to make the cycles unequal
by introducing randomness in this regular dynamo model, as we discuss in the following sections.

\blue{In most of our calculations, we have followed Chatterjee, Nandy, and Choudhuri
(2004) in assuming the meridional circulation to consist of one cell.  Of late,
this assumption has been questioned, although the exact nature of the meridional
circulation in the deeper layers of the convection zone is still not known. We have
shown in Section~4.4 that we can retain the attractive features of our results with
more complicated meridional circulation (Hazra, Karak, and Choudhuri 2014). We have
also included the near-surface shear layer in the calculations presented in Section~4.4.}

\section{Results of Theoretical Modeling}
\subsection{Fluctuations in the Poloidal Field Generation}

We have discussed in the Introduction that the Sun does not produce equal amount of poloidal field 
at the end of every cycle and that the generation of the poloidal field involves randomness. 
Therefore, similar to adding stochastic fluctuations in the traditional
mean-field alpha \citep{Chou92}, adding stochastic fluctuations 
in the B-L $\alpha$ has become a standard practice in the flux transport dynamo community
\citep{CD2000, Jiang07, KarakNandy12}. In the present work, first we introduce stochastic noise in the B-L $\alpha$ in the following way:
\begin{equation}
\alpha_0 \rightarrow \alpha_0 +\sigma(t,\tau) \alpha_0',
\end{equation}\\
where $\tau$ is the coherence time during which the fluctuating component
remains constant and $\sigma$ is a uniformly distributed random number in
the interval [-1, 1]. Considering the typical decay time of the active regions by
surface flux transport process, we fix the coherence time within 0.5 -- 2
months. To see a noticeable effect, we add $75\%$ fluctuations in $\alpha$
({\it i.e.}, $\alpha_0'/\alpha_0=0.75$) with coherence time of $1$ month.
From this stochastically forced model we have to calculate a measure of the 
theoretical sunspot number. We consider the magnetic energy density
($B^2$) of toroidal field at latitude $15^{\circ}$ at the base of
the convection zone ($r = 0.7 \Rs$) as a proxy of sunspot number (this 
was done by Charbonneau and Dikpati, 2000). 
Note that absolute value of the theoretical sunspot number does not have any physical meaning.
Therefore, we scale it by an appropriate factor to match it with the observed sunspot number.
From the time series of theoretical sunspot number, we calculate the cycle periods and decay rates 
in the same way as we have done for the observational data.

\begin{figure}
\centering{ 
\includegraphics[width=1.0\textwidth,clip=]{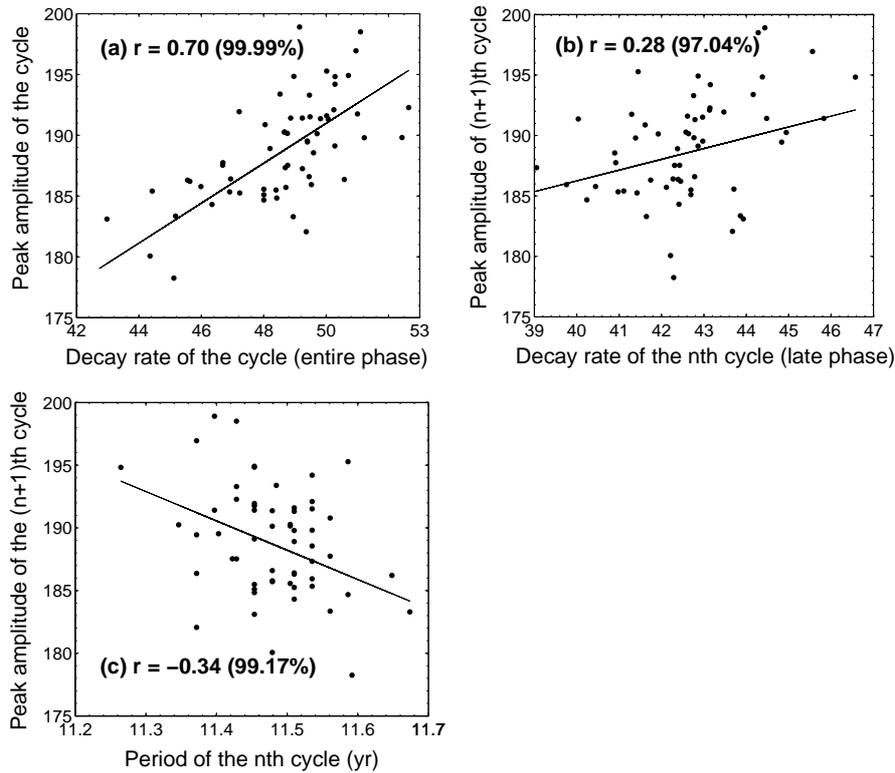}
              }
\caption{Results from stochastically forced dynamo model with B-L $\alpha$ fluctuations: 
Scatter plots showing the correlations between (a) the decay rate and 
the amplitude of cycle $n$, (b) the decay rate of cycle $n$ and the amplitude of 
cycle $n+1$, (c) the period of cycle $n$ and the amplitude of cycle $n+1$.}
\label{alflc}
\end{figure}

In Figure~\ref{alflc}(a) we show the correlation between the decay rates  and 
the amplitudes of the same cycles. We see a positive correlation as in the
observed data presented in Figure~1. It is easy to understand the reason behind getting this positive correlation.
Since we have kept \mc\ fixed, the periods of the solar cycle do not vary much 
but the cycle strengths do vary due to the fluctuations in the poloidal field generation.
Therefore, when the amplitude of a cycle increases while its period remains approximately fixed,
the cycle has to decay rapidly. Hence we find that the stronger cycles decay faster than 
the weaker cycles, producing the positive correlation seen in Figure~\ref{alflc}(a).
However, we see in Figure~\ref{alflc}(b)  that there is not much correlation 
between the decay rate of the cycle $n$ and the amplitude of the next cycle $n+1$
and we are unable to explain the observed correlation seen in Figure~2. 
Note that for Figure~\ref{alflc}(a) the decay rates are calculated from the entire 
decaying part of the cycle which is more appropriate definition of the decay rate as we 
argued in Section~2, whereas for Figure~\ref{alflc}(b) it is computed at the late decay phase 
because observationally we find strong correlation when decay rate is computed in late decay phase
only. Finally we see in Figure~\ref{alflc}(c) that in this study the observed anti-correlation between the period of cycle 
$n$ and the amplitude of cycle $n+1$ (shown in \Fig{percorl}) is also not reproduced. Note that the period does not vary too much when the \mc\ is kept constant.

To sum up, when we introduce fluctuations in the poloidal field generation
mechanism, we can explain the observed correlation between the decay rate and the
amplitude of the cycle shown in Figure~1, but we cannot explain the other observed
correlations presented in Figures~2 and 4.

\subsection{Fluctuations in the Meridional Circulation}
\begin{figure}
\centering{ 
\includegraphics[width=1.0\textwidth,clip=]{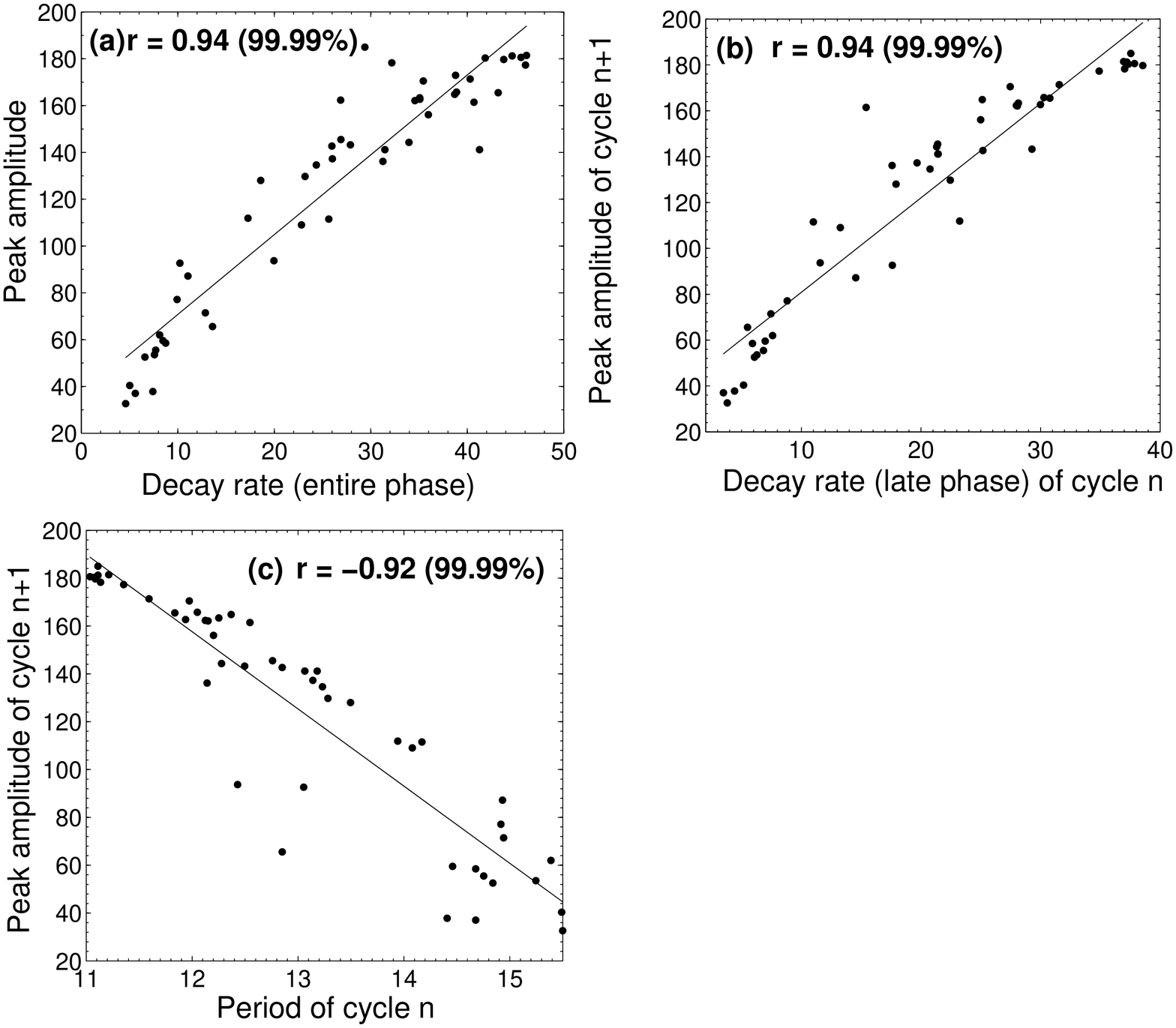}
}
\caption{Same as \Fig{alflc} but with \mc\ fluctuations.}
\label{figmc}
\end{figure}

Next we introduce the other important source of fluctuations in the flux transport dynamo
model, namely, variations of the meridional circulation.
Although we have some observational results of the \mc\ variations near the solar surface for the last 15 -- 20 years,
we do not have long data to conclude the nature of long-term variations
\citep{CD01, Hathaway10b}. However, there are indirect evidences for the
variation of the \mc\ over a long time \citep{LP09, Karak10, PL12}. 
Particularly, Karak and Choudhuri (2011) have used the durations of the past cycles to argue that the \mc\ has long-term variations with the coherence time of probably 20 -- 45~years. There can also be short-term variations
in the \mc\ whose time scale may be related to the convective turnover time of the 
solar convection zone. Such variations with the time scale from a few months to a year are also 
observed in global magnetohydrodynamic simulations \citep{Karak14}.
In this work, we vary the amplitude of the meridional circulation in the same way
as we have done for the $\alpha$ term but with a different coherence
time. We show the results of simulations 
with 30\% fluctuations in the \mc\ with coherence time of 30~years.
\blue{We shall discuss later that various observed correlations can be explained
only if the coherence time is assumed to be not much less than the cycle
period.  While fluctuations of shorter duration (along with spatial variations)
are likely to be present in the meridional circulation, we believe that
they do not play any role in producing the correlations we are studying.} 
With $30\%$ level of fluctuations with a coherence time of 30 years, we get variations
of the \amp\ and of the period in our theoretical model comparable to the observational data. 
As in Section~4.1, we take the time series ($B^2$) 
at latitude $15^{\circ}$ at the base of the convection zone as our proxy
of sunspot activity
and calculate the required correlations from it. 
The relevant correlation plots are shown in Figure~\ref{figmc}. We see in \Fig{figmc}(a) that now the correlation between
the decay rates and the cycle amplitudes has improved. Importantly, the other correlations 
are also correctly reproduced in Figures~\ref{figmc}(b) and \ref{figmc}(c) and can be compared with the
observational plots Figure~2(b) and Figure~4. These correlations did not appear at all when the 
fluctuations in poloidal field generation was introduced ({\it cf.}\ Figures~\ref{alflc}(b) and \ref{alflc}(c)).
To show how the correlations change on changing the correlation time or the level
of fluctuations, we tabulate the values of correlations coefficients under different
situations in Table~2. Each correlation coefficient is calculated from a run of
50 cycles.  It should be kept in mind that there is some statistical noise in 
the values of correlation coefficients. If the correlation coefficient for exactly
the same set of parameters is calculated from different independent runs, the values
for different runs will be a little bit different. Keeping this in mind, we note
that there is no clear trend of the correlation coefficients increasing or decreasing
with increasing levels of fluctuations (other things being the same). However,
all the correlation coefficients tend to decrease on decreasing the coherence time.  

It is not difficult to understand how the correlation in \Fig{figmc}(a) arises.
For a stronger cycle, the sunspot number has to decrease by a larger amount
during the decay phase, making the decay rate faster. However, to understand
the physical reason behind the other two correlations seen in Figures~\ref{figmc}(b) and \ref{figmc}(c),
more subtle arguments are needed. \citet{KarakChou11} extended the
arguments of \citet{Yeates08} and pointed out that a weaker \mc, which
makes the cycles longer, will have two effects. Firstly, the differential
rotation has more time to generate more toroidal field and tends to make the cycles stronger.
Secondly, the turbulent diffusivity gets more time to act on the fields
and tends to make the cycles weaker.
When the diffusivity is high (as in our model), the second effect dominates over
the first and the longer cycles are weaker (the opposite is true for
dynamo models with low diffusivity). \citet{KarakChou11} showed that
this led to an explanation of the Waldmeier effect for dynamo models with
high diffusivity.  We now point out that this tendency (longer cycles tending
to be weaker) is also crucial in our understanding of the correlations seen
in Figures~\ref{figmc}(b) and \ref{figmc}(c).

If the \mc\ keeps fluctuating with a coherence time of 30 years, it would
happen very often that the \mc\ would have a certain value during a cycle (say
cycle $n$) and the early rising phase of the next cycle (say cycle $n+1$). This
is less likely to happen when the coherence time is reduced. Suppose the \mc\
is weaker during cycle $n$ and the rising phase of cycle $n+1$.  Then cycle $n$
will tend to be longer and to have a weaker decay rate.  The following cycle
$n+1$ will have a tendency of being weaker.  This will produce the correlations
seen in Figures~\ref{figmc}(b) and \ref{figmc}(c). On decreasing the coherence time, it will happen less
often that the \mc\ will be the same during cycle $n$ and the rising phase of
the next cycle $n+1$.  Hence the correlations degrade on decreasing the coherence
time of the \mc.

\begin{figure}
\centering{ 
 	\includegraphics[width=1.0\textwidth,clip=]{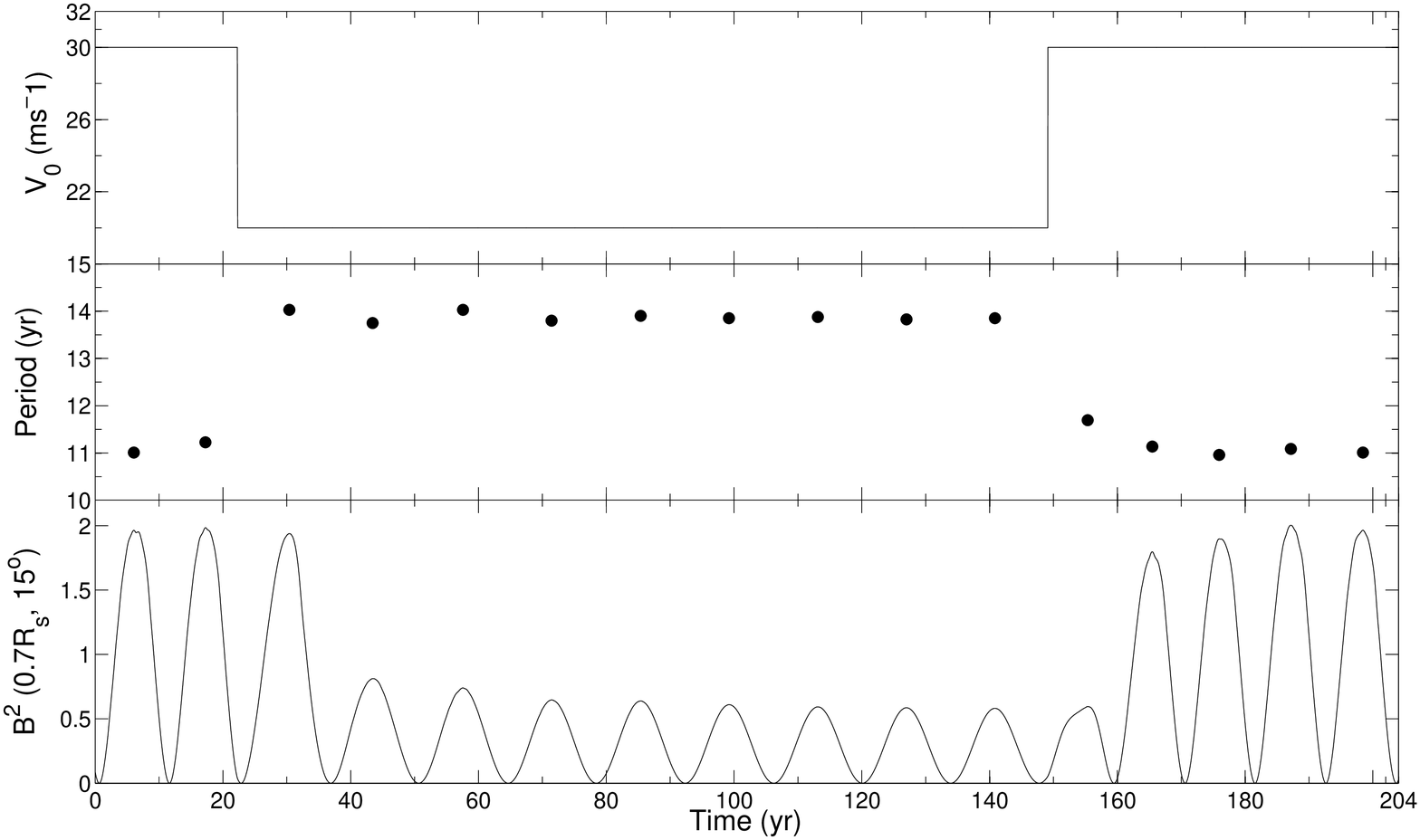}
              }
\caption{Plots showing how the variation of meridional circulation, measured by $v_0$, 
with time (upper panel) changes the period of the cycle (middle panel) and strength of the magnetic field 
(shown by $B^2$ in the lower panel).}
\label{memory}
\end{figure}

We have realized that there is also a memory effect, which enhances the
correlations explained in the previous paragraph. To illustrate this memory
effect, we make a run of our dynamo code in which the \mc\ is decreased
suddenly during a sunspot minimum and then brought back to its original
value during another sunspot minimum a few cycles later.  The \mc\ and
the resulting sunspot activity are plotted in \Fig{memory}.  The periods of
successive cycles are also indicated in the middle panel of \Fig{memory}. On
changing the \mc, it is found that the periods of cycles begin changing
almost immediately.  However, there seems to be a memory effect as far as
the amplitudes of the cycles are concerned.  Even after the \mc\ changes,
the amplitude of the next cycle is very similar to the amplitude corresponding
to the earlier value of the \mc.  This memory effect will certainly enhance
the correlations we are discussing.  Suppose the \mc\ is weaker during the
cycle $n$, making its period longer and decay rate weaker.  Even if the \mc\
becomes stronger by the rising phase of the next cycle $n+1$, the
memory effect will ensure the amplitude of the cycle $n+1$ will still be
weak, thereby producing the correlation.

At this point, we would like to mention a misconception behind the correlation between
cycle $n$ period and cycle $n+1$ amplitude. It may be thought that the overlap between 
two cycles during solar minimum is the cause of this correlation. If the next cycle is 
stronger, then it starts early and the overlap with the present cycle is more. This 
makes the present cycle shorter. However we believe that this is not the source of 
 this correlation because if this is so, then we would have seen this correlation 
in Figure~\ref{alflc}(c) also, where cycle strengths were varied by fluctuations in the 
poloidal field generation. So the overlap is not the reason behind this correlation 
and we only get this in high diffusivity dynamo model with fluctuating \mc.
\begin{table}[h!]
\caption{Correlation coefficients at different levels of fluctuations and coherence time of meridional circulation.}
\begin{tabular}{ccccc}
\hline
&&\multicolumn{2}{c}{Correlation of decay rate}&Correlation of\\
&&\multicolumn{2}{c}{with cycle amplitude of}&previous cycle \\
\cline{3-4}
Coherence time & Fluctuations& Same cycle & Next cycle& period with \\
(year)&($\%$)&(Entire phase)& (Late phase)& amplitude\\ 
\hline
& 10 & 0.92 & 0.92 & -0.97\\
& 20 & 0.86 & 0.92 & -0.95 \\
30 & 30 & 0.87 & 0.89 & -0.96 \\
& 40 & 0.92 & 0.96 & -0.73 \\
& 50 & 0.87 & 0.91 & -0.94\\
\hline
& 10 & 0.79 & 0.85 & -0.95\\
& 20 & 0.86 & 0.86 & -0.98\\
20 & 30 & 0.93 & 0.96 & -0.97 \\
& 40 & 0.90 & 0.87 & -0.88\\
& 50 & 0.89 & 0.90 & -0.97\\
\hline
& 10 & 0.78 & 0.74 & -0.90\\
& 20 & 0.88 & 0.77 & -0.97\\
11 & 30 & 0.90 & 0.85 & -0.92 \\
& 40 & 0.82 & 0.74 & -0.89\\
& 50 & 0.82 & 0.84 & -0.83 \\
\hline
& 10 & 0.70 & 0.63 & -0.87\\
& 20 & 0.83 & 0.74 & -0.86\\
5.5 & 30 & 0.81 & 0.79 & -0.84 \\
& 40 & 0.81 & 0.57 & -0.85\\
& 50 & 0.80 & 0.81 & -0.78\\
\hline
& 10 & 0.57 & 0.48 & -0.78\\
& 20 & 0.58 & 0.59 & -0.64\\
1 & 30 & 0.61 & 0.67 & -0.80 \\
& 40 & 0.73 & 0.25 & -0.65\\
& 50 & 0.69 & 0.38 & -0.72\\
& 75 & 0.64 & 0.39 & -0.58\\
& 100 & 0.65 & 0.73 & -0.76\\
\hline
& 10 & 0.42 & 0.62 & -0.80\\
& 20 & 0.56 & 0.69 & -0.78\\
0.5 & 30 & 0.68 & 0.47 & -0.74 \\
& 40 & 0.62 & 0.56 & -0.67\\
& 50 & 0.61 & 0.56 & -0.79\\
& 75 & 0.64 & 0.50 & -0.81\\
& 100 & 0.64 & 0.60 & -0.87\\
\hline
\label{tabmc}
\end{tabular}
\end{table}

\subsection{Fluctuations in the Poloidal Field Generation and the Meridional Circulation}
\begin{figure}
\centering{ 
\includegraphics[width=1.1\textwidth,clip=]{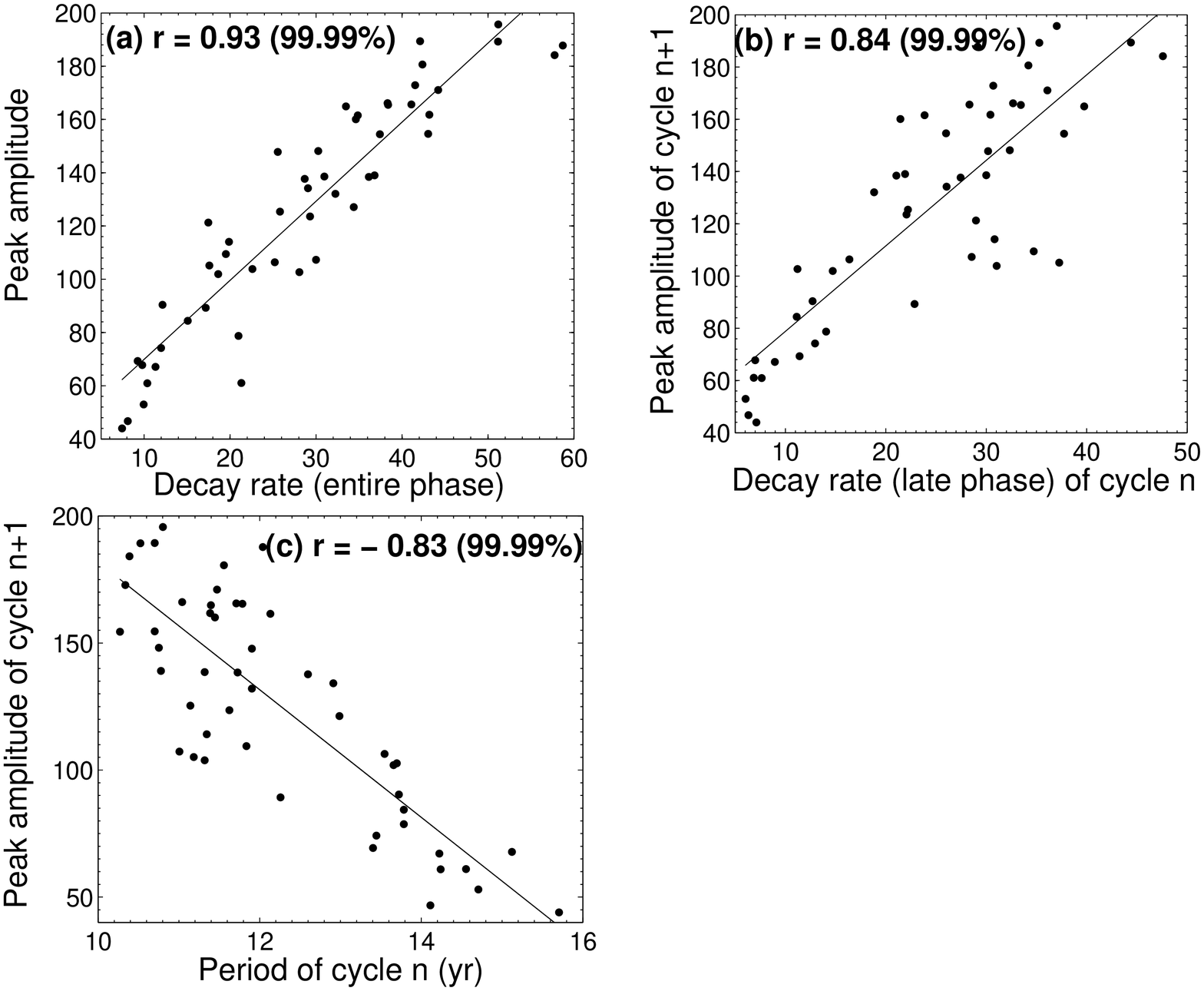}
              }
\caption{Same as \Fig{alflc} but with both B-L $\alpha$ and \mc\ fluctuations.}
\label{figalmc}
\end{figure}

Finally we add fluctuations in both the poloidal field generation process and the meridional 
circulation of the regular model, which is the realistic scenario.
We add the same amount of fluctuations in poloidal field generation
and in meridional circulation that we had added earlier in the individual cases
({\it i.e.}, 75\% fluctuations in the poloidal field generation with
a coherence time of 1 month and 30\% fluctuations in the
meridional circulation with a coherence time of 30~years).
The results are shown in Figures~\ref{figalmc}.
In this figure, we see that the scatters in the correlation plots are very close to 
what we find in actual observations. It is perhaps not a big surprise that all the correlations are reproduced correctly,
because they were already reproduced on introducing fluctuations in \mc\ alone.

\begin{figure}
\centering{ 
\includegraphics[width=1.0\textwidth,clip=]{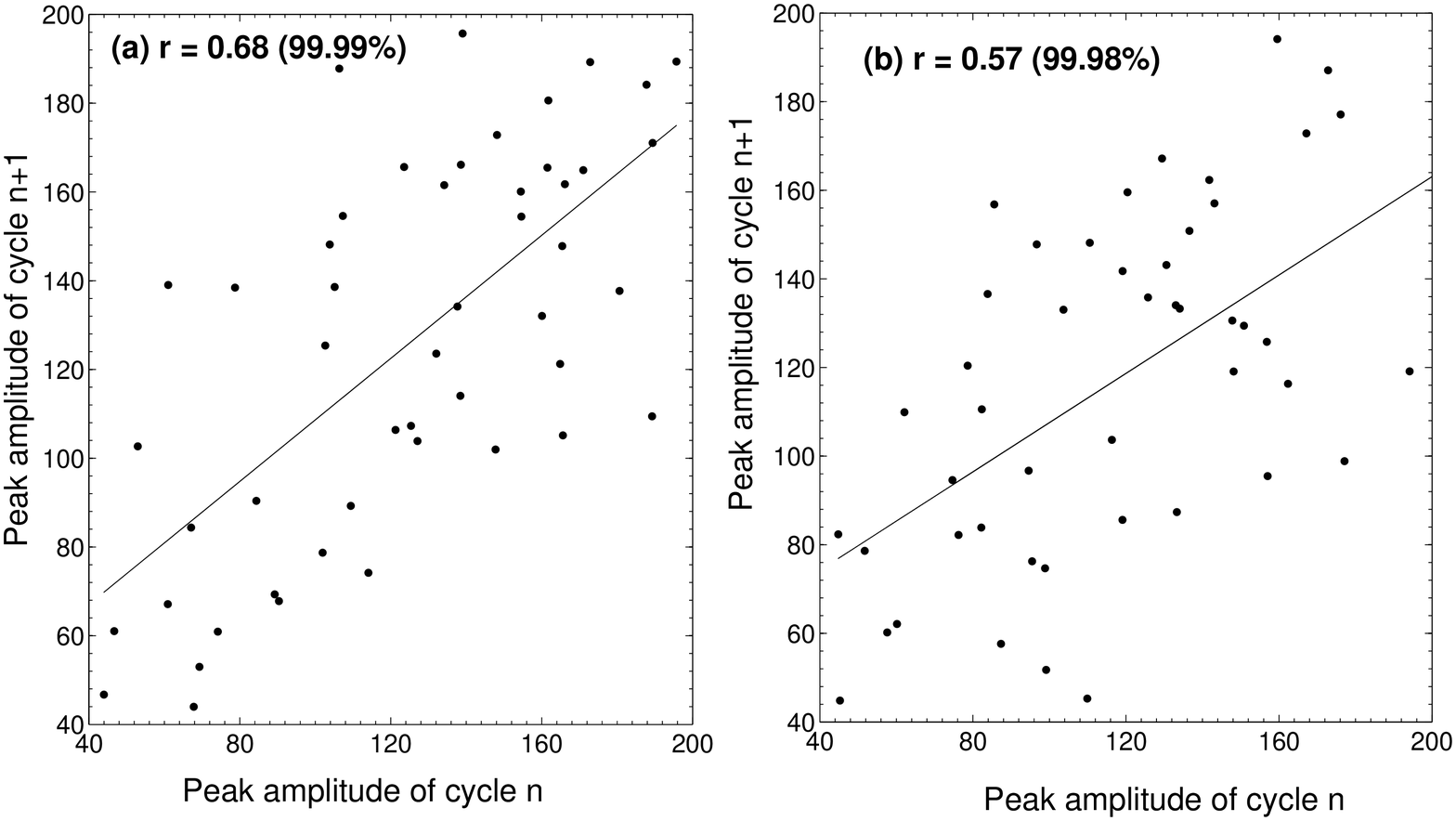}
              }
\caption{(a) Scatter plot of the amplitude of cycle $n$ with the amplitude of cycle $n+1$ with 75\% fluctuation in B-L $\alpha$. (b) Same as (a) but with 100\% fluctuation in B-L $\alpha$.}
\label{theoampl}
\end{figure}

A correct theoretical model also should explain the lack of correlation seen in
Figure~3 between peaks of two successive cycles. \Fig{theoampl}(a) shows the 
correlation between the amplitude of cycle $n$ and the amplitude of cycle $n+1$
for the same level of fluctuations which were used to generate \Fig{figalmc},
whereas \Fig{theoampl}(b) gives the same correlation when the fluctuation is B-L
$\alpha$ is raised to 100\% from 75\%. It is seen that the correlations between
these amplitudes is weak and becomes weaker still on increasing the fluctuation
in the B-L $\alpha$. A physical interpretation is not difficult to give. A
coherence time of 30 years in \mc\ implies that very often the \mc\ will
be the same during two successive cycles, trying to produce a correlation
between the cycles. On the other hand, a fluctuation in the B-L $\alpha$ 
will definitely try to reduce the correlation. Certainly
this fluctuation would try to reduce the correlations seen in \Fig{figalmc} as well.
However, for our choice of parameters, we are able to theoretically
reproduce the three observed correlations as seen in \Fig{figalmc}, whereas
the correlation between successive cycles is much weaker in conformity
with observations.
\blue{We may mention that we also get an anti-correlation between the amplitude
of a cycle and its duration. Our theoretical correlation coefficient ($r = -0.65$)
is somewhat stronger than what Charbonneau and Dikpati (2000) obtained from
the observational data ($r = -0.37$).} 

\subsection{\blue{Robustness of the Results on Changing the  Meridional Circulation and Differential Rotation Profiles}}
\blue{
 So far, our earlier computations are performed using a single-cell meridional circulation 
in each hemisphere. However, recent observations, helioseismic inversions and convection 
simulations suggest the possibility that
the meridional circulation may have a complicated multi-cellular structure rather
than being single-cellular
({\it{e.g.}}, Zhao {\it{et al.,}}\ 2013; Karak {\it{et al.,}}\ 2015). In Hazra, Karak, and Choudhuri (2014), we have shown that
the flux transport dynamo model can reproduce most of basic features of solar cycle 
using multi-structured meridional circulation as long as there is an equator-ward 
flow near the bottom of the convection zone. Therefore we are curious to know whether the 
correlations studied in this paper are also reproduced with multi-structured circulation.
To answer this question, we perform a simulation with three radially stacked 
circulation cells, exactly the same as used in Section~3 of Hazra, Karak, and Choudhuri (2014). For the differential rotation in all our previous works,
we have used a simplified profile of the observed differential rotation 
that does not capture the near-surface shear layer (see {\it e.g.} Figure~1 of Chatterjee {\it et al.,}\ 2004). 
Although it is expected that the near-surface shear layer does not produce significant effect 
on global large-scale fields in the flux transport dynamo \citep{Dikpati02}, 
just for the sake of completeness we use a somewhat improved profile
of differential rotation captured by the following analytical formula}
\begin{eqnarray}
\blue{\Omega(r,\theta) = \sum_{j=0}^2 \cos\left(2j\left(\frac{\pi}{2}-\theta\right)\right)\,\sum_{i=0}^4 c_{ij} (r/R_\odot)^i}.
\end{eqnarray}
\blue{For the coefficients $c_{ij}$ see Table 1 of \citet{Belvedere00}, (see also their Figure~1 for the comparison with observed profile).}

\blue{With these new profiles of the meridional circulation and the differential rotation, we perform 
a dynamo simulation by adding the same amount of stochastic fluctuations 
in B-L $\alpha$ and in meridional circulation as done in the previous section. 
In the results presented earlier, magnetic buoyancy was treated by moving
a part of the toroidal magnetic field to the surface whenever it became larger
than a critical value.  However, as pointed out in \citet{HKC14} and \citet{KKC14},
this way of treating magnetic buoyancy
is not very robust under a large change in parameters and model ingredients.
Therefore, for the computations of this section we use the `non-local' 
magnetic buoyancy as used in Charbonneau and Dikpati (2000), and in many other works. }

\begin{figure}
\centering{ 
\includegraphics[width=1.0\textwidth,clip=]{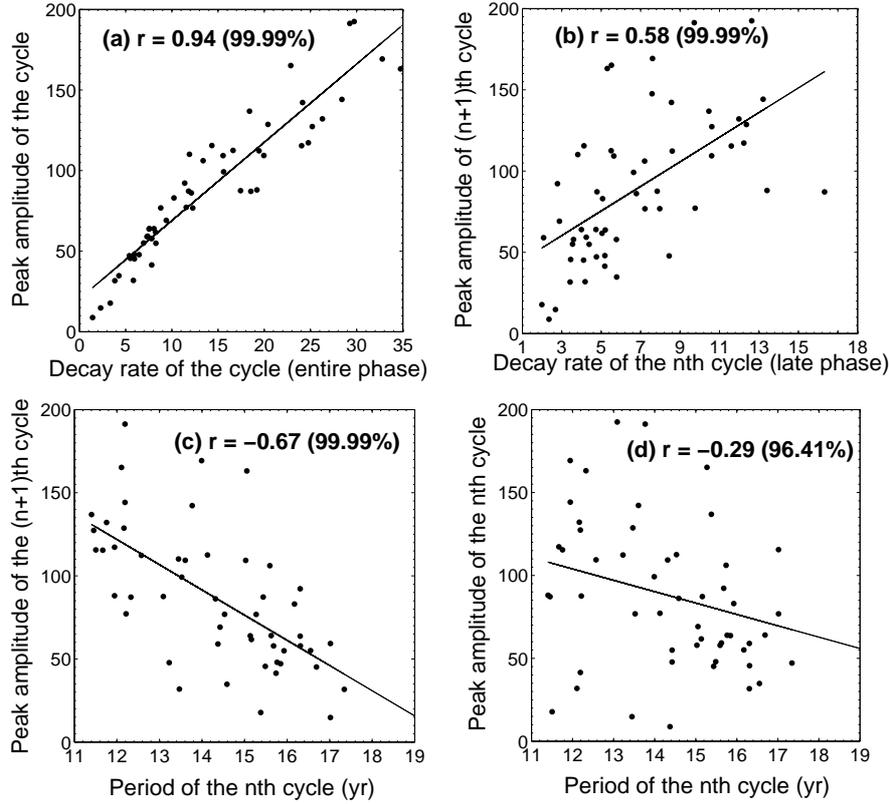}}
\caption{\blue{Same as \Fig{figalmc} but in this model, the large-scale flow has three cells radially stacked in the solar 
convection zone and the differential rotation includes near surface shear layer.}
}
\label{figc3mc}
\end{figure} 

\blue{
Final results from this computation  are shown in Figure \ref{figc3mc}. We observe that even with
the {\it unconventional} meridional flow profile (three radial cells) and addition of near-surface shear layer, 
the correlations do not disappear. Although the correlations in Figures~\ref{figc3mc}(b) and \ref{figc3mc}(c) 
become a little weaker compared to what we have found for the usual single-cell circulation (Figure \ref{figalmc}), 
they show the correct general features as found in observations. The values of the correlations might be
improved a little bit by tuning the amount of imposed fluctuations; we do not want to do that here,
rather we have used same amount of fluctuations as we used in earlier sections.} 

\blue{We make a few remarks about the two ways of treating magnetic buoyancy.
The behaviour of the dynamo can become substantially different on treating magnetic
buoyancy in these two different ways \citep{CNC05}. Since some magnetic
field is removed due to magnetic buoyancy, one would expect the strength of
the toroidal field at the bottom of the convection zone to be depleted due to
the action of magnetic buoyancy.  One unphysical aspect
of the non-local treatment of magnetic buoyancy is that this effect is usually
not taken into account. As we have repeatedly pointed out, one requirement for
obtaining the Waldmeier effect as well as the correlations discussed in this
paper is that the effect of diffusivity has to be more important than the effect
of toroidal field generation. Since the first method of treating magnetic buoyancy
(used in the earlier subsections) puts a cap on the strength of the toroidal field
but not the second non-local method, toroidal field generation remains unrealistically
strong in the second method and it is more difficult to get the correlations
properly in this method. We have taken the magnetic energy density ($B^2$)
at latitude $15^{\circ}$ at the bottom of the convection zone as the proxy of the
sunspot number. In the first method of treating magnetic buoyancy (with the single-cell
meridional circulation, as presented in Section~4.1--3), we found that
all the correlations come out robustly if we use the magnetic energy density ($B^2$)
in a wide range of latitudes as a proxy of the sunspot number. However, on using
the second method of non-local magnetic buoyancy, we find that the magnetic energy
density ($B^2$) has to be taken in a narrow band of low latitudes, with the
correlations disappearing or even reversing if we use the magnetic energy density at higher
latitudes.  To sum up, the second non-local method of treating magnetic buoyancy
is a more robust method and keeps the dynamo stable over a wide range of
parameters (which is not the case with the first method). However, it is more
difficult to reproduce various observed correlations of the solar cycle with this
non-local buoyancy method because the depletion of magnetic field due to buoyancy
is not included.}

\section{Conclusion}
We have discussed three important features of solar cycle -- {i})
a linear correlation between the amplitude of cycle and its decay rate,
{ii}) a linear correlation between the amplitude of cycle $n$ and the decay rate
of cycle $n-1$ and {iii}) an anti-correlation between the amplitude of
cycle $n$ and the period of cycle $n-1$.
We have seen that all these correlations exist in all the data sets considered here.
Last two correlations involve characteristics of one cycle and the amplitude
of the next.  So they provide useful precursors for predicting a future cycle. Just by measuring the
period and the decay rate of a cycle, we can get an idea of the strength of the next cycle. 

We have also explored whether these features can be explained in a B-L type
flux transport dynamo model. We have first introduced stochastic fluctuations in the poloidal
field generation (B-L $\alpha$ term) and we find that only the correlation between the
decay rate and the cycle amplitude is reproduced. However when we added fluctuations
in the \mc, we found that all three correlations are reproduced in qualitative
agreement with observational data.
In our high diffusivity dynamo model, strong \mc\ makes the
period shorter and the decay rate faster, but it also makes the next cycle 
stronger---especially because the cycle strength displays a memory effect,
depending on the \mc\ a few years earlier.
The opposite case happens when \mc\ becomes weaker.
Therefore the fluctuations in the \mc\ are essential to reproduce the observed features.
This study is consistent with earlier studies for modeling the cycle durations and strengths 
of observed cycles \citep{Karak10}, the Waldmeier effect 
\citep{KarakChou11}, grand minima \citep{CK12} and few others
\citep{Passos12} 
which indicate that the variable meridional circulation is crucial in modeling many
aspects of the solar cycle.
\blue{We have found that the observed correlations are reproduced even when
the meridional circulation is assumed to be more complicated than the one-cell pattern
used in most flux transport dynamo models. However, the coherence time of the
fluctuations in the meridional circulation has to be not less than the cycle
period in order to produce the correlations.  The correlations disappear on
making the coherence time too short, implying that fluctuations in the meridional
circulation having coherence time of the order of convective turnover time cannot
be the cause of the observed correlations. The theory of meridional circulation
is still very poorly understood and we have no understanding of what may cause
the fluctuations in meridional circulation with long coherence time.  However,
the pattern in the periods of the past cycles indicate the presence of such fluctuations
(Karak and Choudhuri 2011) and the fact that only such fluctuations can explain
the various observed correlations of the solar cycle convinces us that such fluctuations
in the meridional circulation with long coherence time must exist.}

We have pointed out that the period or the decay rate of a cycle may be used to predict
the next cycle, since these quantities indicate the strength of the \mc\ which
also determines the amplitude of the next cycle a few years later (due to the memory effect).
It seems that the decay rate during the late phase of the cycle is the most
reliable precursor for the next cycle, as seen in Figure~2(b)---presumably because
the decay rate during this phase is the best indicator of the \mc\ during the
particular interval of time which is most crucial in determining the amplitude of
the next cycle. However, fluctuations in the poloidal field generation process
degrades all the observed correlations.  As a result, even Figure~2(b)---displaying
the correlation between the decay rate during the late phase and the amplitude
of the next cycle---has considerable scatter, limiting our ability to predict
the next cycle in this way.

\section*{Acknowledgment}
This work is partly supported by DST through the J. C. Bose Fellowship awarded to ARC. 
GH thanks CSIR, India for financial support. We thank an anonymous referee for careful 
reading and providing constructive comments that helped to improve the quality of the paper.

\bibliographystyle{spr-mp-sola-cnd}
\bibliography{myref}

\end{article}
\end{document}